\documentclass[singlecolumn,epjc3]{svjour3}
\smartqed  
\RequirePackage{graphicx}
\journalname{Eur. Phys. J. C}

\begin{document}

\title{Spherically symmetric electromagnetic mass models of embedding class one}

\author{S.K. Maurya\thanksref{e1,addr1}
\and
Y.K. Gupta\thanksref{e2,addr2}
\and
Saibal Ray\thanksref{e3,addr3}
\and
Sourav Roy Chowdhury\thanksref{e4,addr4}.}

\thankstext{e1}{e-mail: sunil@unizwa.edu.om}
\thankstext{e2}{e-mail: kumar001947@gmail.com}
\thankstext{e3}{e-mail: saibal@iucaa.ernet.in}
\thankstext{e4}{e-mail: b.itsme88@gmail.com}

\institute{Department of Mathematical \& Physical Sciences,
College of Arts \& Science, University of Nizwa, Nizwa, Sultanate
of Oman\label{addr1}
\and
Department of Mathematics, Jaypee
Institute of Information Technology University, Sector-128 Noida,
Uttar Pradesh, India\label{addr2}
\and
Department of Physics,
Government College of Engineering \& Ceramic Technology, Kolkata
700010, West Bengal, India\label{addr3}
\and
Department of
Physics, Seth Anandaram Jaipuria College, Raja Naba Krishna Street,
Kolkata 700005, West Bengal, India\label{addr4}}

\date{Received: date / Accepted: date}

\maketitle

\begin{abstract}
In this article we consider the static spherically symmetric spacetime metric
of embedding class one. Specifically three new electromagnetic mass models are
derived where the solutions are entirely dependent on the electromagnetic field,
such that the physical parameters, like density, pressure etc. do vanish for the
vanishing charge. We have analyzed schematically all these three sets of solutions
related to electromagnetic mass models by plotting graphs and shown that they can
pass through all the physical tests performed by us. To validate these special
type of solutions related to electromagnetic mass models a comparison has been done
with that of compact stars and shown exclusively the feasibility of the models.
\end{abstract}

\keywords{General Relativity; equation of state; electromagnetic mass; compact stars}

\maketitle

\section{Introduction}

It is a widely accepted concept that the $n$ dimensional manifold $V_n$
can be embedded in a pseudo-Euclidean space of $m = n (n+1)/2$
dimensions. The minimum extra dimensions, $m-n = n(n-1)/2$ of the
pseudo-Euclidean space needed is called the embedding class of
$V_n$. In case of the $4$ dimensional relativistic spacetime, the
embedding class is obviously $6$. The well-known cosmological metric of
Friedmann–Lema{\^i}tre–Robertson–Walker (FLRW) \cite{Robertson1935} is of
class $1$, whereas the Schwarzschild's interior and exterior
solutions are of class $1$ and $2$ respectively. The Kerr spacetime
metric has been shown to be of class $5$ \cite{Kuzeev1980}. However,
in the present paper we are limiting ourselves to the static spherically
symmetric metric of embedding class 1 spacetime.

It is seen that the above mentioned metric is compatible only with
two perfect fluid distributions, viz. (i) Schwarzchild's solution~\cite{Schwarzschild1916}
and (ii) Kohler and Chao~\cite{Kohler1965} solution. We would like to exploit
this metric to construct {\it electromagnetic mass} models under the
Einstein-Maxwell framework by considering a charged perfect
fluid distribution. In general, when the charge is zero in a charged
distribution of matter, the subsequent distribution becomes the neutral
counterpart of the charged distribution. This neutral counter
part may belongs to the type of either Schwarzschild interior solution~\cite{Schwarzschild1916}
or Kohler-Chao interior solution~\cite{Kohler1965}.

However, every charged fluid distribution indeed does not possess its neutral
counter part and consequently if the charge is set to zero then the describing
metric turns out to be flat and the corresponding energy density and fluid pressure
will vanish identically. This special type of charged fluid distribution is said to
provide an electromagnetic mass model. In connection with his model
for extended electron Lorentz \cite{Lorentz1904} conjectured that
"there is no other, no 'true' or 'material' mass," and thus proposed
'electromagnetic masses of the electron'. Later on Wheeler \cite{Wheeler1962}
and Wilczek \cite{Wilczek1999} pointed out that electron has a "mass
without mass". Feynman, Leighton and Sands \cite{Feynman1964} actually termed this type
of models as "electromagnetic mass models". For further reading on historical notes
and technical works on the electromagnetic mass one may look at the Ref. \cite{Ray2008}
and \cite{Florides1962,Cooperstock1978,Tiwari1984,Gautreau1985,Gron1985,Tiwari1986,Leon1987,Tiwari1991,Ray2004,Ray2006}
respectively under the framework of Einstein-Maxwell theory.

Unfortunately, the electromagnetic mass models proposed by most of the above investigators \cite{Cooperstock1978,Tiwari1984,Gautreau1985,Gron1985,Tiwari1986,Leon1987,Tiwari1991,Ray2004,Ray2006}
suffer from a negative pressure or density of the fluid due to the equation of state of the form
\begin{equation}
\rho + p = 0, \label{eos}
\end{equation}
where $\rho$ is the density and $p$ is the pressure. This type of equation of state
in the literature known as a 'false vacuum' or 'degenerate vacuum' or '$\rho$-vacuum'~\cite{Davies1984,Blome1984,Hogan1984,Kaiser1984}.
It has been argued that though, in general, this equation of state leads to negative pressure but provides easier
junction conditions and realistic expression for mass~\cite{Tiwari1984,Gautreau1985,Gron1985,Bonnor1989}.
Although the junction conditions do not require the density to vanish at the boundary
as is true for gaseous spheres. Such a model is available in the literature for
both uncharged and charged cases~\cite{Mehra1966,Mehra1980}. However, we also note that
the classical models of electron should contain the regions of negative density~\cite{Kuchowicz1968,Krori1975}.
It would be interesting to mention that a Weyl-type character of the field has been
attributed which form electromagnetic mass model~\cite{Cooperstock1989}.

In the present study we have attempted to obtain a charged fluid of class $1$
by choosing specific metric potential(s) of the class $1$ such
that they do not form a sub-set of the metric potentials of
Schwarzschild's interior metric (inclusive de-Sitter and Einstein
universe) and Kohler-Chao metric~\cite{Kohler1965}. We argue here that
the static spherically symmetric metric of embedding class one is more
suitable to construct electromagnetic mass model as it possess
lesser number of neutral counterparts of the charged fluids in
comparison to general static spherical symmetric metric. Now if the charge
be zero in the charged fluid, the describing metric will turn into flat by virtue
of the structure of the metric. In the past, several alternatives
were used by several investigators to obtain the electromagnetic mass
models~\cite{Tiwari1984,Gautreau1985,Gron1985,Tiwari1986,Leon1987}
by employing the equation of state (\ref{eos}) as pure charge condition~\cite{Gautreau1985}
which takes the equivalent form as $g_{11}g_{44} = -1$~\cite{Tiwari1984}. On the other hand,
Ponce de Leon \cite{Leon1987} has utilized the charged Einstein's clusters \cite{Florides1974,Gron1986}
to get the electromagnetic mass models. For further studies on different aspects of 
electromagnetic mass models one may look at the Refs.~\cite{Ray1993,Tiwari1997,Ray2002,Ray2007,Das2011,Usmani2011,Rahaman2012a}.

However, for the construction of electromagnetic mass models we invoke
a different method by adopting an algorithm which is very efficient
to generate solutions of the desired form and physics, as such no
{\it ad hoc} assumptions are required to obtain electromagnetic mass models.
The main motivation of the present paper, therefore, is to obtain a set of solutions for
the electromagnetic mass model with the help of charged fluid distribution
of spherically symmetric class one metric. The logic behind considering
the class one metric is that if one removes charge from the solutions then
either the Schwarzschild solution~\cite{Schwarzschild1916} or
Kohler-Chao solution~\cite{Kohler1965} will emerge the metric being flat
and all the physical parameters - pressure, density etc. - become zero.
To this aim our scheme of investigations are as follows: in Sect. 2
we provide the class one metric and fit the metric potentials in to the
Einstein-Maxwell field equations for the spherically symmetric matter
distribution. The next part is to construct electromagnetic mass
models for stellar systems we provide the necessary algorithm (Sect. 3)
and by exploiting the mathematical formalism we generate three new set
of solutions in connection with electromagnetic mass models (Sect. 4).
We also discuss the boundary conditions regarding all these solutions
and determine the unknown constants of integration (Sect. 6). As till now
we don't know exact nature of the solutions set so we adopt in Sect. 7
some specific techniques to explore the different features and properties
of the electromagnetic mass models for physical acceptability of the
anisotropic stellar models. In Sect. 8 we try to validate the solutions set
related to electromagnetic mass models with some of the observed compact star
candidates. We discuss our results in the concluding Sect. 9.

\section{The class one metric and the Einstein-Maxwell field equations}

Let us consider the static spherical symmetric metric to be
\begin{equation}
ds^2 = - e^{\lambda} dr^2 - r^2(d\theta^2 + sin^2\theta d\phi^2) + e^{\nu} dt^2, \label{metric1}
\end{equation}
which may represent the space time of emending class $1$,
if it satisfies the Karmarker condition \cite{Karmarkar1948}
\begin{equation}
R_{1414}=\frac{R_{1212}R_{3434} + R_{1224}R_{1334}}{R_{2323}}, \label{2}
\end{equation}
with $R_{2323} \neq 0$ \cite{Pandey1982}.

The above condition along with (\ref{metric1}) yields the following differential equation
\begin{equation}
\frac{\lambda' \nu'}{\left(1-e^\lambda\right)}=-2\left( \nu''+\nu'^2 \right)+\nu'^2+\lambda' \nu', \label{3}
\end{equation}
with the constraint $e^\lambda\neq 1$, where $\lambda$ and $\nu$ are metric potentials
of the line element (\ref{metric1}) which are function of radial coordinate $r$ only.

The solution of the above differential equation (\ref{3}) can be obtained as
\begin{equation}
e^\lambda=\left(1 + K \frac{\nu'^2 e^\nu}{4}\right), \label{4}
\end{equation}
where $K$ is non zero arbitrary constant, $\nu'(r)\neq 0$,
$e^{\lambda(0)}=1$ and $\nu'(0)=0$.

If Eqs. (\ref{metric1}) and (\ref{4}) describe a charge perfect fluid
distribution then the functions $\lambda(r)$ and $\nu(r)$ must
satisfy the Einstein-Maxwell field equations
\begin{equation}
{G^i}_j =  {R^i}_j - \frac{1}{2} R {g^i}_j = \kappa ({T^i}_j + {E^i}_j), \label{5}
\end{equation}
where $\kappa = 8\pi$ is the Einstein constant with $G=c=1$
in the relativistic geometrized units.

The matter within the star is assumed to be locally a perfect
fluid and consequently $T^i_j$ and $E^i_j$, the energy-momentum
tensors for the fluid distribution and the electromagnetic field tensors,
are respectively defined by
\begin{equation}
{T^i}_j = [(c^2 \rho + p)v^iv_j - p{\delta^i}_j, \label{6}
\end{equation}

\begin{equation}
{E^i}_j = \frac{1}{4 \pi}(-F^{im}F_{jm} + \frac{1}{4}{\delta^i}_jF^{mn}F_{mn}),\label{7}
\end{equation}
where $v^i$ is the four-velocity as
$e^{\lambda(r)/2}v^i={\delta^i}_4$, $\theta^i$ is the unit space-like vector
in the direction of radial vector, $\theta^i = e^{\lambda(r)/2}{\delta^i}_1$, $\rho$ is the matter-energy density and
$p$ is the fluid pressure.

The above anti-symmetric electromagnetic field tensor $F_{ij}$ in Eq. (\ref{7}),
denotes the velocity and can be defined
\begin{equation}
F_{ij} = \frac{\partial A_j}{\partial A_i} - \frac{\partial A_i}{\partial A_j},
\end{equation}
This should satisfy the Maxwell equations
\begin{equation}
F_{ik,j} + F_{kj,i} + F_{ji,k} = 0,
\end{equation}
and
\begin{equation}
\frac{\partial}{\partial x^k}({\sqrt -g} F^{ik}) = -4\pi{\sqrt -g}
J^i, \label{8}
\end{equation}
where $g$ is the determinant of quantities $g_{ij}$ in Eq. (\ref{8}) and is given by
\begin{center}
$g = \left(\begin{array}{cccc} e^{\nu} & 0           & 0    & 0\\
                              0       &-e^{\lambda} & 0    & 0\\
                              0       & 0           & -r^2 & 0\\
                              0       & 0           & 0    & -r^2sin^2\theta \end{array} \right) = - e^{(\nu+\lambda)}r^4 sin^2\theta $,

\end{center}
where $A_j=(\phi(r), 0, 0, 0)$ is the four-potential and $J^i$ is
the four-current vector defined by
\begin{center}
$J^i = \frac{\sigma}{\sqrt g_{00}} \frac{dx^i}{dx^0} = \sigma v^i,$
\end{center}
where $\sigma$ is the charged density.

For static matter distribution the only non-zero component of the
four-current is $J^4$ and because of the spherical symmetry this has
only a functional relation with the radial coordinate $r$.
The only non-vanishing component of the electromagnetic field tensor
($F^{41} = - F^{14}$) describes the radial component of the
electric field. Hence, from Eq. (\ref{8}), one can easily get the expression
for the electric field
\begin{equation}
F^{41} = e^{-(\nu+\lambda)/2} \left[\frac{q(r)}{r^2}\right],
\end{equation}
where $q(r)$ represents the electric charge contained within the
sphere of radius $r$ and is defined by
\begin{equation}
q(r) = 4\pi \int_0^r \sigma r^2 e^{\lambda/2} dr = r^2 \sqrt{-F_{14}F^{14}} = r^2 F^{41} e^{(\nu+\lambda)/2}.\label{9}
\end{equation}

Equation (\ref{9}) can be treated as the relativistic version of Gauss's law which reduces to the following form:
\begin{equation}
\frac{\partial}{\partial r}(r^2 F^{41} e^{(\nu+\lambda)/2}) = - 4\pi r^2 e^{(\nu+\lambda)/2} J^4.
\end{equation}

For the spherically symmetric metric (\ref{metric1}), the
Einstein-Maxwell field equations can be expressed in the following
ordinary differential equations
\begin{equation}
-\kappa {T^1}_1 = \frac{{\nu}^{\prime}}{r} e^{-\lambda} - \frac{(1 - e^{-\lambda})}{r^2} = \kappa p - \frac{q^2}{r^4},\label{e1}
\end{equation}

\begin{equation}
-\kappa {T^2}_2 = -\kappa {T^3}_3 = \left[\frac{{\nu}^{\prime\prime}}{2} - \frac{{\lambda}^{\prime}{\nu}^{\prime}}{4} + \frac{{{\nu}^{\prime}}^2}{4}
+ \frac{{\nu}^{\prime} - {\lambda}^{\prime}}{2r}\right]e^{-\lambda} = \kappa p + \frac{q^2}{r^4},\label{e2}
\end{equation}

\begin{equation}
-\kappa {T^4}_4 = \frac{{\lambda}^{\prime}}{r} e^{-\lambda} + \frac{(1 - e^{-\lambda})}{r^2} = \kappa \rho + \frac{q^2}{r^4},\label{e3}
\end{equation}
where the prime denotes differentiation with respect to the radial coordinate $r$.

By using the Eqs. (\ref{e1}), (\ref{e2}), (\ref{e3}) and also (\ref{4}), we obtain
\begin{equation}
\frac{\nu'}{r^2 \left(4 + K \nu'^2 e^\nu\right)}(4r-K \nu')=\kappa p - \frac{q^2}{r^4}, \label{e4}
\end{equation}

\begin{equation}
 \frac{4}{\left(4 + K \nu'^2 e^\nu\right)}\left[\frac{\nu'}{2r} - \frac{\left( K \nu' e^\nu-2r \right) \left(2\nu''+\nu'^2 \right)}{2r \left(4 + K \nu'^2 e^\nu\right)} \right]= \kappa p + \frac{q^2}{r^4}, \label{e5}
\end{equation}

\begin{equation}
\frac{K \nu' e^\nu}{\left(4 + K \nu'^2 e^\nu\right)}\left[\frac{4(2\nu''+\nu'^2)}{\left(4 + K \nu'^2 e^\nu\right)}+\frac{\nu'}{r}\right]  = \kappa \rho + \frac{q^2}{r^4}.\label{e6}
\end{equation}

On the other hand, the pressure isotropy condition can be given by
\begin{equation}
\left(\frac{k\nu' e^\nu}{2r}-1\right) \left[\frac{2\nu'}{r\left(4
+ K \nu'^2 e^\nu\right)}-\frac{4(2\nu''+\nu'^2)}{\left(4 + K
\nu'^2 e^\nu\right)^2}\right]=\frac{2q^2}{r^4}. \label{e7}
\end{equation}

A closer observation of the above set of differential equations
easily indicates that if charge vanishes in a charged
fluid of embedding class one, then survived neutral counterpart
will only be either the Schwarzschild~\cite{Schwarzschild1916} interior solution
(or its special cases de-sitter universe or Einstein's universe) or the
Kohler-Chao~\cite{Kohler1965} solution, otherwise either the charge cannot be zero
or the survived space-time metric will become flat.

Now, one can look at Eq. (\ref{e7}) which immediately indicates that
in absence of the charge either of the two factors on the left hand
side has to be zero. Consequently, it can be shown that if the
first factor of  Eq. (\ref{e7}) be zero then it gives rise to
the Kohler-Chao~\cite{Kohler1965} solution in the form:
\begin{equation}
ds^2 = -\frac{(A+2Br^2)}{(A+Br^2)} dr^2 - r^2(d\theta^2 +
sin^2\theta d\phi^2) + (A+Br^2) dt^2. \label{e8}
\end{equation}

The pressure and density, in this model, are
\begin{equation}
\kappa p = \frac{B}{(A+2Br^2)}, \label{e9}
\end{equation}

\begin{equation}
\kappa \rho = B\frac{(3A+2Br^2)}{(A+Br^2)}. \label{e10}
\end{equation}

One can observe from (\ref{e9}) that since it does not possess zero pressure
as well as density for any finite radius on the surface, it
cannot represent a compact star.

Let us now consider the second factor of Eq. (\ref{e7}) which in its vanishing form
ultimately provides the Schwarzschild~\cite{Schwarzschild1916} interior solution
\begin{equation}
ds^2 = -\left(1-\frac{r^2}{R^2}\right)^{-1} -r^2(d\theta^2 +
sin^2\theta d\phi^2) + \left( A+B\sqrt{1-\frac{r^2}{R^2}} \right)
^2 dt^2, \label{e11}
\end{equation}
with its pressure and density as follows:
\begin{equation}
\kappa p = - \frac{A+3B\sqrt{1-\frac{r^2}{R^2}
}}{R^2\left(A+B\sqrt{1-\frac{r^2}{R^2} }\right) }, \label{e12}
\end{equation}

\begin{equation}
\kappa \rho = \frac{3}{R^2}, \label{e13}
\end{equation}
where $A$ and $R$ are non-zero constant quantities and $B > 0$.

If the mass function for electrically charged fluid sphere is
denoted by $m(r)$, then it can be defined in terms of the metric
function $e^{\lambda(r)}$ as
\begin{equation}
e^{-\lambda(r)} = 1 - \frac{2m(r)}{r} + \frac{q^2}{r^2}, \label{e14}
\end{equation}
where the function $m(r)$ represents the gravitational mass of the matter
contained in a sphere of radius $r$. Now, if $R$ represents the radius
of the fluid sphere then it can be shown that $m$ is a constant with
$m(r = R) = M$ outside the fluid distribution where $M$ is the
gravitational mass. Following the work of Florides~\cite{Florides1974} this
can be defined as
\begin{equation}
M = \mu(R) + \xi(R),\label{mass}
\end{equation}
where $\mu(R) = \frac{\kappa}{2} \int_0^R\rho r^2 dr$ is the mass
inside the sphere, $\xi(R) = \frac{\kappa}{2} \int_0^R \sigma r q
e^{\lambda/2}dr$ is the mass equivalence of the electromagnetic
energy of distribution and $Q = q(R)$ is the total charge inside
the fluid sphere.

By using Eq. (\ref{mass}) one can write the mass, in terms of
energy density and charge function, as follows:
\begin{equation}
m(r)= \frac{\kappa}{2} \int \rho r^2 dr +\frac{1}{2} \int
\frac{q^2}{r^2}dr + \frac{q^2}{2r},\label{e15}
\end{equation}
Again from Eqs. (\ref{e1}) and (\ref{e4}) we obtain expression for metric potential
\begin{equation}
\nu^{\prime} = \frac{\left(\kappa r p + \frac{2m}{r^2} -
\frac{2q^2}{r^3}\right)}{\left(1 - \frac{2m}{r} +
\frac{q^2}{r^2}\right)}.\label{e16}
\end{equation}

Also, the expression for the pressure, in its gradient form, can be obtained
by using Eqs. (\ref{e1}) and (\ref{e4}) - (\ref{e6}) as follows
\begin{equation}
\frac{dp}{dr} = -\frac{M_G(r)\left(p +
\rho\right)}{r^2}e^{\left(\lambda-\nu\right)/2} + \frac{q}{4 \pi
r^4}\frac{dq}{dr}, \label{e17}
\end{equation}
where $M_G$ is the gravitational mass within the sphere of radius $r$ and is
given by
\begin{equation}
M_G(r)=\frac{1}{2}r^2\nu'e^{\left(\nu-\lambda\right)/2}.
\end{equation}

The above Eq. (\ref{e17}) represents the charged generalization of the
Tolman-Oppenheimer-Volkoff (TOV) equation of continuity for perfect fluid
stellar system~\cite{Tolman1939,Oppenheimer1939}.

\section{Algorithm for electromagnetic mass models}

We are now in position to construct `electromagnetic mass' models of stellar system
for class one metric by using algorithm given by Maurya et al. \cite{Maurya2015}.

The Eqs. (\ref{e1}) - (\ref{e3}) in terms of mass function reduce to
\begin{equation}
-\frac{2m(1+r\nu')}{r^3} + \frac{\nu'}{r}+\frac{q^2(1+r\nu')}{r^4}+ \frac{q^2}{r^4}=\kappa p,\label{e18}
\end{equation}

\begin{eqnarray}
-\frac{m'(2+r{\nu'})}{2r^2} - \frac{m(2r^2\nu'' +r^2{\nu'}^2 +r\nu' -2)}{2r^3} \nonumber\\
 + \frac{2rqq'\nu'-2q^2\nu'+4qq' +(r^2+q^2)(2r\nu'' +r\nu'^2+2\nu)}{4r^3}
- \frac{2q^2}{r^4}=\kappa p ,\label{e19}
\end{eqnarray}

\begin{equation}
\frac{2m'}{r^2} - \frac{2qq'}{r^3}= \kappa  c^2 \rho. \label{e20}
\end{equation}

From Eqs. (\ref{e18}) and (\ref{e19}), the first order linear
differential equation for $m(r)$ in terms of $\nu(r)$ and electric
charge function $q(r)$ can be provided as follows:
\begin{equation}
m'+ \frac{(2r^2\nu'' + r^2\nu'^2 - 3\nu'r -6)}{r(r\nu' + 2)}m
= \frac{r(2r\nu'' + r\nu'^2 - 2\nu')}{2(r\nu' + 2)} + f(r),\label{e21}
\end{equation}
where
\begin{equation}
f(r)=\frac{q^2\left[ 2r^2\nu'' + r\nu'(r\nu' - 4)
-16\right]}{2r^2(r\nu' + 2)} + \frac{qq'(r\nu' + 2)}{r(r\nu' + 2)}.
\label{e22}
\end{equation}

Hence the mass function $m(r)$ can be given by
\begin{equation}
m(r)= e^{-\int g(r) dr} \left[\int\left\{h(r) +
f(r)\right\}\left(e^{\int g(r)dr}\right) dr + A
\right],\label{e23}
\end{equation}
where
\begin{equation}
g(r) = \frac{(2r^2\nu'' + r^2\nu'^2 - 3r\nu' -6)}{r(r\nu' + 2)},
\end{equation}
and
\begin{equation}
h(r) = \frac{r(2r^2\nu'' + r\nu'^2 - 2\nu'}{2(r\nu' + 2)}. \label{e24}
\end{equation}

\section{New class of electromagnetic mass models for stellar systems}

\subsection{Solution of Type I}
We consider the following suitable function
\begin{equation}
\nu(r) = 2Ar^2+ logB, \label{e25}
\end{equation}

\begin{equation}
\lambda(r) = log\left(1 + K \frac{\nu'^2 e^\nu}{4}\right),
\end{equation}
where $A$ and $B$ are positive constants.

The expressions for the mass and the electric charge are respectively
\begin{equation}
\frac{2m(r)}{r} =
Ar^2\left[\frac{De^{2Ar^2}}{1+DAr^2e^{2Ar^2}}+\frac{Ar^2(D^2e^{4Ar^2}
+ 4 - 4De^{2Ar^2})}{2(1+DAr^2e^{2Ar^2})^2}\right],\label{e26}
\end{equation}

\begin{equation}
\frac{q^2}{r^4} = E^2 = A^2r^2\left[\frac{D^2e^{2Ar^2}
+4-4De^{2Ar^2}}{2(1+DAr^2e^{2Ar^2})^2}\right], \label{e27}
\end{equation}
where $D=4ABK$ is a pure constant.

Again, the expression for the energy density and the pressure are given by
\begin{equation}
8 \pi \rho =A\left[ \frac{D^2Ar^2e^{4Ar^2}
-4Ar^2+6De^{2Ar^2}(2Ar^2+1)}{2(1+DAr^2e^{2Ar^2})^2}\right], \label{e28}
\end{equation}

\begin{equation}
8 \pi p =A\left[
\frac{-D^2Ar^2e^{4Ar^2}+4(2+Ar^2)+2De^{2Ar^2}(2Ar^2-1)}{2(1+DAr^2e^{2Ar^2})^2}\right].\label{e29}
\end{equation}

The respective gradients of above physical parameters are
\begin{equation}
\frac{dp}{dr} =
-\frac{2A^2r}{8\pi}\left[\frac{-D^3Ar^2e^{6Ar^2}+D^2e^{4Ar^2}(-3+4Ar^2+8A^2r^4)+4De^{2Ar^2}(4+7Ar^2+4A^2r^4)-4}{2(1+DAr^2e^{2Ar^2})^3}\right], \label{e30}
\end{equation}

\begin{equation}
\frac{d\rho}{dr} =
-\frac{2A^2r}{8\pi}\left[\frac{D^3Ar^2e^{6Ar^2}+D^2e^{4Ar^2}(11+20Ar^2+28A^2r^4)-4De^{2Ar^2}(6+7Ar^2+4A^2r^4)+4}{2(1+DAr^2e^{2Ar^2})^3}\right]. \label{e31}
\end{equation}

\subsection{Solution of Type II}
Here the functional relation for the metric potentials are
\begin{equation}
\nu(r) = 2log (1+ \sinh Ar^2)+ logB, \label{e32}
\end{equation}

\begin{equation}
e^{\lambda(r)} = \left(1 + K \frac{\nu'^2 e^\nu}{4}\right),
\end{equation}
where $A$ and $B$ are positive constants.

The expressions of mass and electric charge are
\begin{eqnarray}
\frac{2m(r)}{r} = Ar^2\frac{D\cosh^2 Ar^2}{(1+DAr^2\cosh^2 Ar^2)}
\nonumber\\ -Ar^2\left(\frac{D(2\sinh Ar^2 \cosh Ar^2(1+\sinh
Ar^2)+2\cosh^3 Ar^2)-D^2\cosh^4 Ar^2(1+\sinh Ar^2) -4\sinh
Ar^2}{2(1+DAr^2\cosh^2 Ar^2)^2(1+\sinh Ar^2)}\right),\label{e33}
\end{eqnarray}

\begin{eqnarray}
\frac{q^2}{r^4} = E^2 = A^2r^2\left(\frac{-D\left[(2\sinh Ar^2 \cosh Ar^2-D\cosh^4
Ar^2)(1+\sinh Ar^2)+2\cosh^3 Ar^2\right] +4\sinh
Ar^2}{2(1+DAr^2\cosh^2 Ar^2)^2(1+\sinh Ar^2)}\right).\label{e34}
\end{eqnarray}

The energy density and pressure (taking $ x= Cr^2 $) are given by
\begin{equation}
8 \pi \rho = A\frac{-4x \sinh x + D\left[2x\cosh^3 x +(6\cosh^2
x+10x\sinh x \cosh x +Dx\cosh^4 x)(1+\sinh
x)\right]}{2(1+Dx\cosh^2 x)^2(1+\sinh x)}, \label{e35}
\end{equation}

\begin{equation}
8 \pi p = A\frac{8 \cosh x +4x \sinh x + D\left[6x\cosh^3 x
-(2\cosh^2 x+2x\sinh x \cosh x +Dx\cosh^4 x)(1+\sinh
x)\right]}{2(1+Dx\cosh^2 x)^2(1+\sinh x)}. \label{e36}
\end{equation}

As done in the previous case, the expressions for the pressure
and the density gradient can be determined by taking their derivatives
with respect to $r$ which are not produced here being their
very complicated forms.

\begin{figure}[h]
\centering
\includegraphics[width=0.5\textwidth]{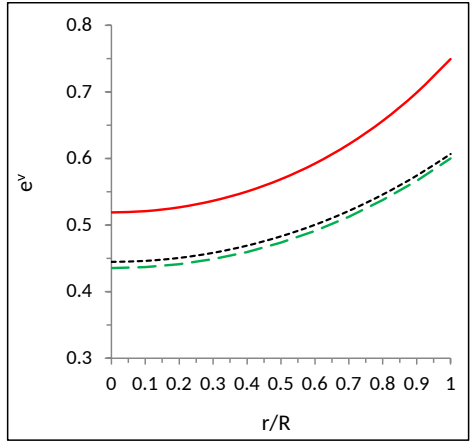}
\caption{$e^{\nu}$ are plotted with continuous line for solution I,
small dashed line for solution II and long dashed line for solution III.}
\end{figure}

\begin{figure}[h]
\centering
\includegraphics[width=0.5\textwidth]{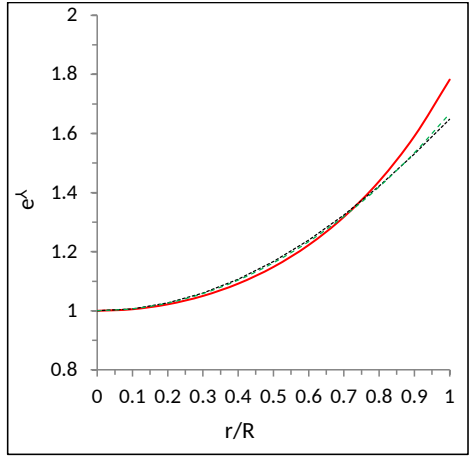}
\caption{$e^{\lambda}$ are plotted with continuous line for solution I,
small dashed line for solution II and long dashed line for solution III.}
\end{figure}

\subsection{Solution of Type III}
The metric potentials in this case are related to the
following functions
\begin{equation}
\nu(r) = 2log (1+ \sin Ar^2)+ logB, \label{e37}
\end{equation}

\begin{equation}
e^{\lambda(r)} = \left(1 + K \frac{\nu'^2 e^\nu}{4}, \right)
\end{equation}
where $A$ and $B$ are positive constant.

The expressions of mass and electric charge are
\begin{eqnarray}
\frac{2m(r)}{r} = Ar^2\frac{D\cos^2 Ar^2}{(1+DAr^2\cos^2 Ar^2)}
\nonumber\\ +Ar^2\left(\frac{D(\sin 2Ar^2(1+\sin Ar^2)-2\cos^3
Ar^2)+D^2\cos^4 Ar^2(1+\sin Ar^2) -4\sin Ar^2}{2(1+DAr^2\cos^2
Ar^2)^2(1+\sin Ar^2)}\right),\label{e38}
\end{eqnarray}

\begin{eqnarray}
\frac{q^2}{r^4} = E^2 = A^2r^2\left(\frac{D\left[\sin 2Ar^2 (1+\sin Ar^2)-2\cos^3
Ar^2\right]+D^2\cos^4 Ar^2(1+\sin Ar^2) -4\sin
Ar^2}{2(1+DAr^2\cos^2 Ar^2)^2(1+\sin Ar^2)}\right). \label{e39}
\end{eqnarray}

The expression for energy density and pressure (taking $ x= Cr^2$) are respectively
\begin{equation}
8 \pi \rho = A\frac{4x \sin x + D\left[2x\cos^3 x +(6\cos^2
x-5x\sin x )(1+\sin x)\right]+D^2x\cos^4 x(1+\sin x)}{2(1+Dx\cos^2
x)^2(1+\sin x)}, \label{e40}
\end{equation}

\begin{equation}
8 \pi p = A\frac{8 \cos x -4x \sin x + D\left[6x\cos^3 x -(2\cos^2
x-x\sin 2x +Dx\cos^4 x)(1+\sin x)\right]}{2(1+Dx\cos^2 x)^2(1+\sin
x)}, \label{e41}
\end{equation}

Likewise the case II, the expressions for the pressure
and the density gradients being very cumbersome we are leaving
those calculations of respective derivatives.

Let us look at Figs. 1 and 2 regarding the desirable features
on the basis of their respective solution. It is expected that
the solution should be free from physical and geometrical
singularities, i.e. the fluid pressure and the energy density at
the center should be finite and metric potentials $e^{\lambda(r)}$
and $e^{\nu(r)}$ should have non-zero positive values in the range
$0 \leq r \leq R$. At the center one must have $e^{\lambda(0)}=1$ and
$e^{\nu(0)}=B$ for each solution. Interestingly, both Figs. 1 and 2
show that metric potentials are positive and finite at the center.

Similarly, the density $\rho$ should be positive and the pressure
$p$ must be positive inside the star as well as it should be zero
at the boundary of the fluid sphere. All these features are quite
available from Figs. 3 and 4.

\begin{figure}[h]
\centering
\includegraphics[width=0.5\textwidth]{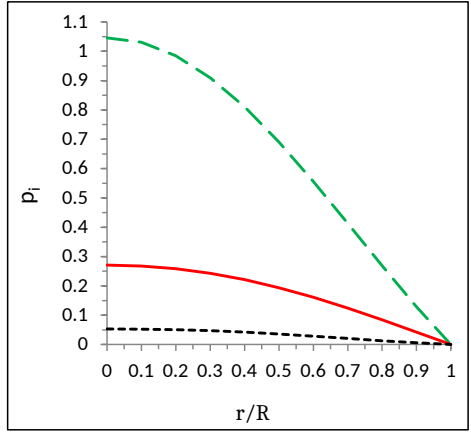}
\caption{ Pressure is plotted with long dashed line for solution-I,
continuous line for solution II and small dashed line for solution III.}
\end{figure}

\begin{figure}[h]
\centering
\includegraphics[width=0.5\textwidth]{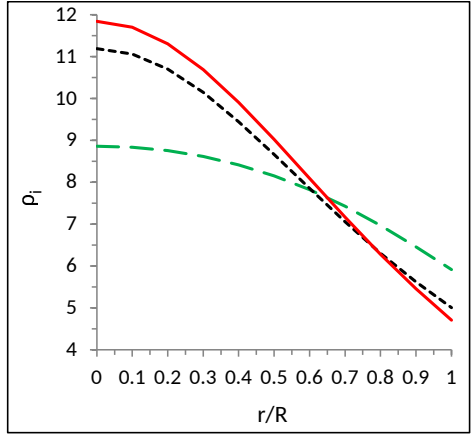}
\caption{Density is plotted with long dashed line for solution I,
small dashed line for solution II, continuous line for solution III.}
\end{figure}

Let us summarize the above results and at a glance try to get the flavor
of these. We would like to mention here that if $A=0$ in the cases (I), (II)
and (III) then the corresponding metrics at once turn to flat spacetime
and also the expressions for the electric charge, the pressure and the energy density
automatically vanish. Therefore, the three charged fluid distributions obtained
above depict the three electromagnetic mass models.

\section{Boundary conditions for the spherical system}

The above system of equations has to be solved under the condition
that the radial pressure $p=0$ at $r=a$ (where $r=a$ is the outer
boundary of the fluid sphere). The interior metric (\ref{metric1}) can join smoothly
at the surface of spheres to the Reissner-Nordstr{\"o}m metric~\cite{Misner1964}
\begin{equation}
ds^2=-\left(1-\frac{2M}{r}+\frac{Q^2}{r^2}\right)^{-1}dr^2
-r^2(d\theta^2 +\sin^2 \theta
d\phi^2)+\left(1-\frac{2M}{r}+\frac{Q^2}{r^2}\right)dt^2. \label{e42}
\end{equation}

This requires the continuity of $e^\lambda(r)$, $e^\nu(r)$ and $q(r)$
across the boundary $r=R$.
\begin{equation}
e^{-\lambda(R)} = \left(1-\frac{2M}{R}+\frac{Q^2}{R^2}\right),
\end{equation}

\begin{equation}
e^{\nu(R)} = \left(1-\frac{2M}{R}+\frac{Q^2}{R^2}\right),
\end{equation}

\begin{equation}
q(R)=Q,
\end{equation}

\begin{equation}
p_{(r=R)}=0.
\end{equation}

By using all the above boundary conditions we are able to find out
expressions for various constants as can be seen below.

\subsection{Solution of Type I}

\begin{equation}
D=\frac{(2AR^2-1) {^+_-} \sqrt{(8A^2R^4+4AR^2+1)}}{AR^2e^{2AR^2}}
\end{equation}

At the boundary:
\begin{equation}
e^{-\lambda(R)} =e^{\nu(R)},
\end{equation}
gives
\begin{equation}
B=\frac{1}{e^{2AR^2}(1+DAR^2e^{2AR^2})}.
\end{equation}

\subsection{Solution of Type II}

The pressure is zero on the boundary $r=R$ (here $X=AR^2$)
and we obtain expressions for the constants as follows:
\begin{equation}
D=\frac{D_1(X)^+_-\frac{1}{2}\sqrt{16X\cosh^4 X(1+\sinh X)(2\cosh
X +X\sinh X)-D_2(X)}}{-X\cosh^4 X(1+\sinh X)},
\end{equation}
where
\begin{equation}
D_1(X)= \cosh^2 X(1-3X\cosh X+\sinh X)+X\cosh X \sinh X(1+\sinh X)
\end{equation}
and
\begin{equation}
D_2(X)=-\cosh^2 X\left(4X-2\cosh X+2X\cos 2X-2X\sin X-\sinh 2X\right)^2.
\end{equation}

At the boundary:
\begin{equation}
e^{-\lambda(R)}=e^{\nu(R)},
\end{equation}

gives
\begin{equation}
B=\frac{1}{\left(1+\sinh X\right)^2\left(1+DX\cosh^2 X\right)}.
\end{equation}

\subsection{Solution of Type III}

The pressure being zero on the boundary $r=R$ (here $X=AR^2$)
the constants can be given by
\begin{equation}
D=\frac{\cos^2 X(6X\cos X-2\sin X-2)+X\sin 2X(1+\sin X)}{2X\cos^4 X(1+\sin X)},
\end{equation}

\begin{equation}
\frac{{^+_-}\sqrt{-16X\cos ^4 X(1+\sin X)(-2 \cos X+X\sin
X)+[6X\cos^3 X-(2\cos^2 X-X\sin 2X)(1+\sin X)]^2}}{-2X\cos^4
X(1+\sin X)}.
\end{equation}

At the boundary:
\begin{equation}
e^{-\lambda(R)}=e^{\nu(R)},
\end{equation}
gives
\begin{equation}
B=\frac{1}{\left(1+\sin X\right)^2\left(1+DX\cos^2 X\right)}.
\end{equation}

\section{Physical features of the electromagnetic mass models for stellar systems}
In the solution part (Sect. 4) we have analyzed some of the physical parameters,
potentials, density, pressure etc., through their graphical plots. They exhibited
desirable physical features regarding stellar configuration. However, in the present
Sect. 6 we are interested to perform a few rigorous tests for other physical parameters,
velocity and charge, and also prepare a check list for energy conditions and stability issues
(such as TOV equation and Buchdahl condition).

\subsection{Sound velocity}
The velocity of sound should monotonically decrease away from the center
and increase with the increase of density, i.e. $\frac{d}{dr}\left(\frac{dp}{d\rho}\right) <0$ or
$\frac{d^2p}{d\rho^2}>0$ for $0 \leq r \leq R$. It is argued by Canuto~\cite{Canuto1973}
that the equation of state at ultra-high distribution of matter the sound speed
decreases outwards.

In the present model, from Fig. 5, it is clear that velocity is decreasing for
solution I and increasing for solution II and III throughout the
star. Therefore, the solutions for solution II and III are not suitable
at all as far as compact star is concerned. This is because the equation of
state for nuclear matter shows a regular behavior of
$\frac{dp}{d\rho} $ for these solutions~\cite{Durgapal1964}.

The above discussions, based on the demonstration of the figures, immediately
restraint ourselves to study henceforth only the solution of type I electromagnetic mass model.

\begin{figure}[h]
\centering
\includegraphics[width=0.5\textwidth]{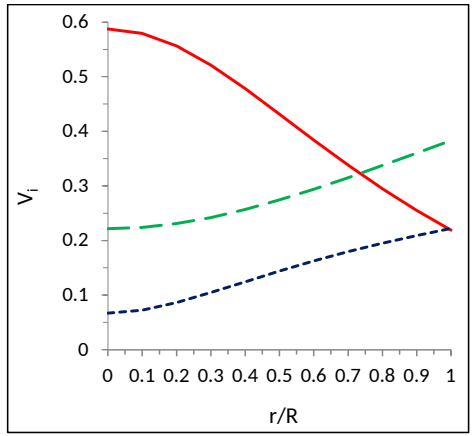}
\caption{Velocity is plotted with continuous line for solution I,
long dashed line for solution II and small dashed line for solution III.}
\end{figure}

\subsection{Electric charge for solution I}

From the present model it is observed that in the unit of Coulomb,
the charge on the boundary is $1.15295 \times 10^{20}$~C and at the
center it is zero (as the charge on the boundary is $0.9889$
so we have to multiply this by the number $1.1659 \times 10^{20}$
to obtain the resultant numerical value).

One can observe from Fig. 6 that the charge profile starts from a minimum
and acquires the maximum value at the boundary. This figure has been drawn
for the compact star $RX~J~1856-37$ with the constant values $CR^2=0.1836$,~$D=2.9540$.

\begin{figure}[h]
\centering
\includegraphics[width=0.5\textwidth]{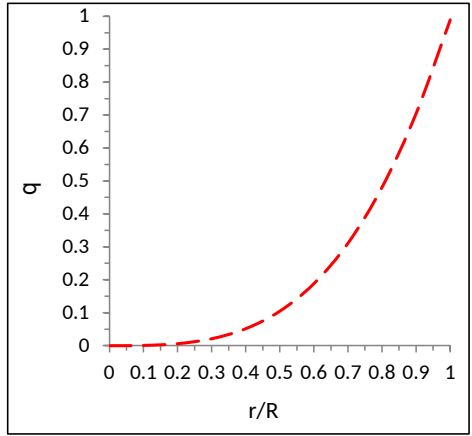}
\caption{Electric Charge is plotted with long dashed line for solution I.  }
\end{figure}

\subsection{Energy Conditions for solution of type I}

For physical validity an energy-momentum tensor has to obey the
following energy conditions:
\begin{enumerate}
\item null energy condition (NEC): $\rho+\frac{E^2}{4\pi} \geq 0$,
\item weak energy condition (WEC): $\rho-p+\frac{E^2}{4\pi} \geq 0$,
\item strong energy condition (SEC): $\rho-3p+\frac{E^2}{4\pi} \geq 0$.
\end{enumerate}

We have plotted the feature of different energy conditions in Fig. 7
for the values of different physical parameters connected to energy conditions
for the constants: $CR^2=0.1836$,~$M=0.9041~M_{\odot}$,~$R=6.006~Km$ and $\frac{M}{R}=0.222$.
The figure indicates that all the energy conditions are satisfied
throughout the interior region of the stellar system.

\begin{figure}[h]
\centering
\includegraphics[width=0.5\textwidth]{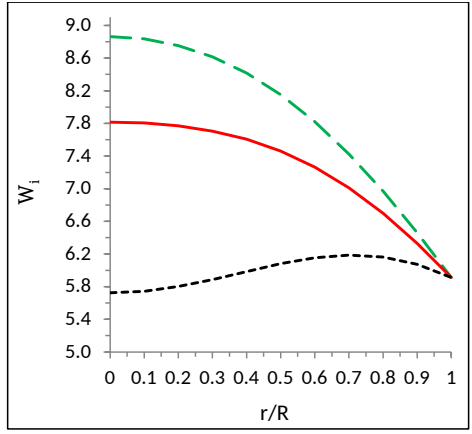}
\caption{NEC is plotted with long dashed line, WEC is plotted with continuous line
and SEC is plotted with small dashed line for solution I.}
\end{figure}

\subsection{Generalized TOV equation for solution of type I}

We write the generalized Tolman-Oppenheimer-Volkoff (TOV) equation~\cite{Varela2010}
in the following form:
\begin{equation}
-\frac{M_G(\rho+p_r)}{r^2}e^{(\lambda-\nu)/2}-\frac{dp}{dr}+
\sigma \frac{q}{r^2}e^{\lambda/2} =0,
\end{equation}
where $M_G$ is the effective gravitational mass within the radius $r$ and can be provided
\begin{equation}
M_G(r)=\frac{1}{2}r^2 \nu^{\prime}e^{(\nu - \lambda)/2}.
\end{equation}

The above TOV equation describes the equilibrium condition for a
charged fluid subject to gravitational ($F_g$), hydrostatic
($F_h$) and electric ($F_e$) forces. Therefore, one can write it in a more suitable form
\begin{equation}
F_g+F_h+F_e=0,
\end{equation}
where
\begin{equation}
F_g=-\frac{1}{2}
\nu'(\rho+p)=-\frac{2A^2r}{8\pi}\left[\frac{2De^{2Ar^2}(1+4Ar^2)+4}{(1+DAr^2e^{2Ar^2})^2}\right],
\end{equation}

\begin{equation}
F_h=-\frac{dp}{dr}
=\frac{2A^2r}{8\pi}\left[\frac{-D^3Ar^2e^{6Ar^2}+D^2e^{4Ar^2}(-3+4Ar^2+8A^2r^4)+4De^{2Ar^2}(4+7Ar^2+4A^2r^4)-4}{2(1+DAr^2e^{2Ar^2})^3}\right],
\end{equation}
and
\begin{equation}
F_e=\sigma
\frac{q}{r^2}e^{\lambda/2}=\frac{A^2r}{4\pi}\left[\frac{(-2+De^{Ar^2})
\left[-6+D^2Ar^2e^{4Ar^2}+De^{2Ar^2}(3+2Ar^2+8A^2r^4)\right]}{2(1+DAr^2e^{2Ar^2})^3}\right],
\end{equation}

The plot for the TOV equation is shown in Fig. 8. We  observe from this figure
that the system under the joint balancing action of the different forces, e.g.
gravitational, hydrostatic and electric to attain an overall static equilibrium.
However, from Fig. 8 it is also clear that the gravitational force has the
dominant role over the hydrostatic force whereas the electric force has the
negligible contribution to the equilibrium. This feature seems quite reasonable
in the case of the compact stellar system.

\begin{figure}[h]
\centering
\includegraphics[width=0.5\textwidth]{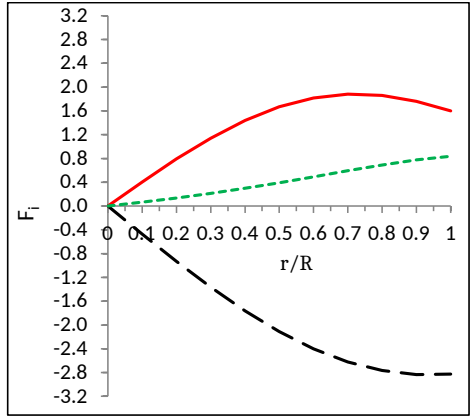}
\caption{$F_g$ is plotted with long dashed line, $F_h$ is plotted with continuous line
and $F_e$ is plotted with small dashed line for solution I.}
\end{figure}

\subsection{Effective mass-radius relation and surface redshift for solution of type I}

Buchdahl \cite{Buchdahl1959} has proposed an absolute constraint of the
maximally allowable mass-to-radius ratio $(M/R)$ for isotropic
fluid spheres in the form $2M/R \leq 8/9$. However, B{\"o}hmer and Harko~\cite{Boehmer2007}
have shown that for a compact object with charge, $Q (<M)$, there is a
lower bound for the mass-radius ratio
\begin{equation}
\frac{3Q^2}{2R^2} \left(\frac{1+\frac{Q^2}{18R^2}}{1+\frac{Q^2}{12R^2}}\right) \leq \frac{2M}{R},
\end{equation}
whereas the upper bound of the mass-radius of a charged sphere was generalized
by Andreasson~\cite{Andreasson2009} as follows:
\begin{equation}
\sqrt{M} \leq \frac{\sqrt{R}}{3} + \sqrt{\frac{R}{9} + \frac{Q^2}{3R}}.
\end{equation}

In the present model, the effective gravitational mass is given by
\begin{equation}
M_{eff}=4\pi\int_{0}^{R}\left(\rho +\frac{E^2}{8\pi}\right) r^2
dr = \frac{1}{2}R[1-e^{-\lambda(R)}]= \frac{1}{2}R\left[\frac{DAR^2e^{2AR^2}}{1+DAR^2e^{2AR^2}}\right].
\end{equation}

Therefore, the compactness factor can be written as
\begin{equation}
u= \frac{M_{eff}}{R}= \frac{1}{2}\left[\frac{DAR^2e^{2AR^2}}{1+DAR^2e^{2AR^2}}\right]. \label{eq35}
\end{equation}

The surface redshift in connection with the above compactness is given by
\begin{equation}
Z = (1-2u)^{-1/2} - 1 = e^{\lambda(R)/2} - 1=\sqrt{1+DAR^2e^{2AR^2}} -1. \label{eq36}
\end{equation}

The plot of the surface redshift is shown in Fig. 9 for the compact star $RX~J~1856-37$
with the constant values $CR^2=0.1836$,~$D=2.9540$. It can be observed that
there is a gradual increase in the redshift which is an acceptable physical feature.
The maximum surface redshift for the present stellar configuration of radius $R = 6.006$~Km
turns out to be $Z = 0.3882$ which seems well within the limit $Z \leq 2$~\cite{Buchdahl1959,Straumann1984,Boehmer2006}.

\begin{figure}[h]
\centering
\includegraphics[width=0.5\textwidth]{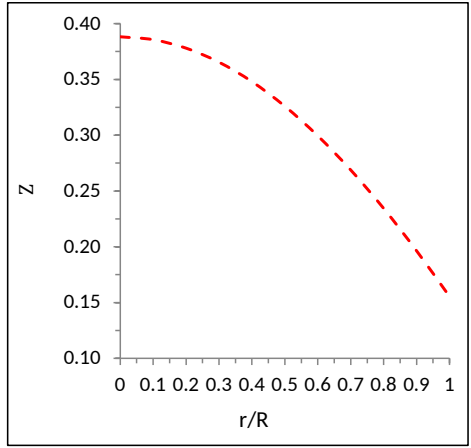}
\caption{Red shift is plotted with long dashed line for solution I.}
\end{figure}

\section{Validating the model with strange star candidates}

In the foregoing Sect. 6 we have studied several physical behavior of stellar system
in connection with electromagnetic mass models. In some of the subsections, e.g.
6.2 (electric charge) and 6.5 (surface redshift), we have also shown
graphical plots specifically for the compact star $RX~J~1856-37$ with the mass 
$M=0.9041~M_{\odot}$ and the radius $R=6.006$~Km.

However, it seems that some more investigations are needed to show the validity
of our models for other compact stars which have definite observed physical features.
In the following two Tables 1 and 2 we, therefore, produce data sheet for the
purpose of comparison between the present model stars and the observed compact stars.

\begin{table}[h]
\centering \caption{Values of the model parameters $A$, $B$, $D$ and $K$
for different strange stars}\label{tbl-4}
\begin{tabular}{@{}lrrrrrrrrrrrrr@{}} \hline
Strange star  &$M~(M_\odot)$ & $R~(Km)$  & $M/R$    & $B$   & $A$                     &$D$     & $K$  \\
candidates \\ \hline

$RX~J~1856-37$ &0.9041     &6.006     &0.222     &0.5189 &$5.0962\times 10^{-13}$  &2.9540  &$2.7925 \times 10^{12}$\\

$Her~X-1$     & 0.9825     & 6.700    & 0.216    &0.5552 &$3.6319\times 10^{-13}$  &3.0626  &$3.7972 \times 10^{12}$\\

$PSR~1937+21$ &2.1         &11.4998   &0.269     &0.4103 &$1.9503\times 10^{-13}$  &2.5857  &$8.0775 \times 10^{12}$\\

$PSRJ ~1614-2230$ &1.97    &11.3664   &0.2553    &0.4419 &$1.8112\times 10^{-13}$  &2.6998  &$8.4338 \times 10^{12}$\\

$PSRJ~0348+0432$  &2.1     &11.7372   &0.2636    &0.4228 &$1.8017\times 10^{-13}$  &2.6315  &$8.6369 \times 10^{12}$\\   \hline
\end{tabular}
\end{table}

\begin{table}[h]
\centering \caption{Energy densities and pressure for different
strange star candidates for the above parameter values of
Table 5}\label{tbl-4}
\begin{tabular}{@{}lrrrrrrrrr@{}}
\hline

Strange star            & Central Density           & Surface density          & Central pressure \\

candidates              & ($gm/cm^{-3}$)            & ($gm/cm^{-3}$)           & ($dyne/cm^{-2}$)\\ \hline

$RXJ~1856-37$           & $2.4252 \times 10^{15}$   & $1.6183  \times 10^{15}$ & $2.2243  \times 10^{35}$\\

$Her~X-1$               & $1.8869  \times 10^{15}$  & $1.2718\times 10^{15}$   & $2.5768 \times 10^{35}$\\

$PSR~1937+21$           & $8.1241  \times 10^{14}$  & $5.0473 \times 10^{14}$  & $1.3334 \times 10^{35}$\\

$PSRJ ~1614-2230$       & $7.4524  \times 10^{14}$  & $4.9876 \times 10^{14}$  & $1.1384 \times 10^{35}$\\

$PSRJ~0348+0432$        & $7.6379  \times 10^{14}$  & $4.7797 \times 10^{14}$  & $1.1918 \times 10^{35}$\\ \hline

\end{tabular}
\end{table}

For our model we particularly note that for the compact star $RXJ~1856-37$ with mass 
$M=0.9041~M_{\odot}$ and radius $R=6.006$~Km the surface redshift turns out to be 
$Z = 0.3882$ which seems falls within the range $Z \leq 2$~\cite{Buchdahl1959,Straumann1984,Boehmer2006} 
and $0 < Z \leq 1$~\cite{Rahaman2012b,Kalam2012,Hossein2012,Kalam2013,Bhar2015}.
However, one may figure out the surface redshifts for other compact stars also as 
provided in Tables 1 and 2 and we expect those values will be within the above specified range.
On the other hand, surface density as can be seen from Table 2 is of the order of $10^{14} - 10^{15}$~gm/cc.
This very high density indicates that the model under `electromagnetic mass' represents 
an ultra-compact star~\cite{Ruderman1972,Glendenning1997,Herjog2011}.

Therefore, we would like to pass a general remark that our models in connection with `electromagnetic mass' 
represent compact stars of several categories.

\section{Conclusion}

We have considered the static spherically symmetric spacetime metric
of embedding class one in the present investigation. Is has been possible 
to show the existence of electromagnetic mass models specifically in 
connection with compact stars. Three new electromagnetic mass models are
derived where the solutions are entirely originating from the electromagnetic field,
such that the density and pressure like physical parameters do vanish for the
vanishing charge alone. However, a meticulous analysis reveals that among these 
three sets of solutions all are not equally interesting as far as astrophysical 
several aspects are concerned. To validate these special type of solutions 
related to electromagnetic mass models, we have also conducted a comparison 
between our proposed model and the observed compact stars which shows satisfactory
results in favor of the present theoretical modeling.

However, an obvious question may arise to the study of the compact stellar 
configuration under Einstein-Maxwell spacetime, especially how the charge 
comes in the consideration of such kind of systems. A brief historical note 
on the issues of stability of static spherically symmetric stellar systems 
and as an effective measure of averting singularity why one should include 
charge and what is the process of holding huge amount of charge inside 
the bodies are available exhaustively in the Ref.~\cite{Ray2007} and 
the Refs. therein. As an continuation of this discussion, we feel, an outline 
on the charged bodies may be helpful to the readers which follows in the next paragraph. 

In the history of general relativity the first ever exact solutions of the 
Einstein field equations, the well-known Schwarzschild interior solutions, 
suffer from the problem of singularity due to gravitational collapsing of a 
spherically symmetric matter distribution. One way to overcome this singularity 
is to include electrical charge to the neutral bodies. It is suggested by the 
scientists that gravitational collapse can be avoided in the presence of charge 
where the gravitational attraction is counter balanced by the electrical repulsion 
in addition to the pressure gradient~\cite{Felice1995,Sharma2001,Ivanov2002}.
To this end questions came up regarding the stability of the charged sphere 
and also about the amount of charge that holds by the star. A good amount of works 
have been done by several authors on the stability issue~\cite{Bonnor1965,Stettner1973,Glazer1976,Glazer1979,Pant1979,Whitman1981}. 
On the other hand, in some recent studies~\cite{Ray2003,Ghezzi2005,Varela2010} 
we find out estimate of electric charge in the compact stars which amounts a 
huge charge of the order of $10^{19} - 10^{20}$~Coulomb.

\section*{Acknowledgments}
SKM acknowledges support from the authority of University of
Nizwa, Nizwa, Sultanate of Oman. Also SR is thankful to
the authority of Inter-University Center for Astronomy and
Astrophysics, Pune, India for providing Associateship
programme under which a part of this work was carried out.

\end{document}